\documentclass[aps,prl,twocolumn,superscriptaddress,reprint]{revtex4-2}

\usepackage{amsmath,amsthm,amsfonts,mathrsfs,bm,bbm,amssymb}	
\usepackage[dvipsnames,usenames]{xcolor}
\usepackage{empheq}
\usepackage[citecolor=black,linkcolor=black]{hyperref}

\newcommand*{\commutator}[2]{\mathinner{[{#1},{#2}]}}

\begin{document}

\title{Pseudomode description of general open quantum system dynamics: non-perturbative master equation for the spin-boson model}

\author{Graeme Pleasance}
\email{gpleasance1@gmail.com}
\affiliation{Quantum Research Group, School of Chemistry and Physics, University of KwaZulu-Natal, Durban, 4001, South Africa}

\author{Francesco Petruccione}
\affiliation{Quantum Research Group, School of Chemistry and Physics, University of KwaZulu-Natal, Durban, 4001, South Africa}
\affiliation{National Institute for Theoretical and Computational Sciences (NITheCS), South Africa}

\date{\today} 

\begin{abstract}

We outline a non-perturbative approach for simulating the behavior of open quantum systems interacting with a bosonic environment defined by a generalized spectral density function. The method is based on replacing the environment by a set of damped harmonic oscillators---the pseudomodes---thereby forming an enlarged open system whose dynamics is governed by a Markovian master equation. Each pseudomode is connected to one of the poles of the spectral density when analytically continued to the lower-half complex frequency plane. Here, we extend previous results to a completely generic class of open system models, and discuss how our framework can be used as a powerful and versatile tool for analyzing non-Markovian open system dynamics. The effectiveness of the method is demonstrated on the spin-boson model by accurately benchmarking its predictions against numerically exact results. 

\end{abstract}

\maketitle

In any realistic setting a quantum system will interact with its surrounding environment, and so the theoretical modelling of its behavior must account for the effects of this interaction \cite{BreuerTOQS2002}. For environments which are structured or strongly coupled to the system, these interactions can induce time-dependent changes in the system state that are strongly characterized by memory, and for these cases the resulting behavior is non-Markovian \cite{Vega2017,Breuer2016,Li2018}. Over recent years the study of non-Markovian systems has seen intense focus not only out of fundamental interest \cite{Wolf2008,Rivas2010,Rivas2014,Breuer2009,Lorenzo2013,Chruscinski2014}, but also due to the increasing ability to probe quantum systems over shorter time and length scales at which memory effects play a heightened role. For example, energy transport in light-harvesting complexes and nanoscale devices \cite{Ishizaki2012,Nalbach2011,Chin2013}, as well as photonic crystals \cite{John1990,John1994,Hoeppe2012}, have been shown to display distinctive features associated with non-Markovian dynamics. Applications of non-Markovianity to quantum metrology \cite{Chin2012,Wu2020} and quantum information processing \cite{Bylicka2014,Huelga2012,Li2020} have too been the subject of much interest.

Various techniques capable of simulating memory effects within the dynamics of open systems have been introduced to date, including the hierarchical equations of motion (HEOM) \cite{Tanimura1989,Ishizaki2005,Tanimura2020}, path integrals \cite{Makri1995,Makri1995a}, and tensor networks \cite{Rosenbach2016,Strathearn2018,Gribben2020}, among others \cite{Hu1992,Zhang2012,Chin2010,Tamascelli2019,White1992}. While a number of these methods rely on keeping track of the environment state explicitly \cite{Chin2010,Tamascelli2019,White1992}, an alternative approach consists of tracing out the environmental degrees of freedom in favor of obtaining a master equation solely for the reduced density matrix of the system \cite{BreuerTOQS2002,Hu1992,Zhang2012}. Such master equations can in principle be derived non-perturbatively through the use of projection operator techniques \cite{Nakajima1958,Zwanzig1960,Shibata1977}. However, their complicated structure often renders an exact description infeasible without the use of certain approximations. The Born-Markov approximations can lead to a much more convenient description in terms of a Markovian master equation either in Bloch-Redfield \cite{Redfield1965} or Gorini-Kossakowski-Sudarshan-Lindblad (GKSL) form \cite{Gorini1976,Lindblad1976}, but this typically comes at the expense of restricting the validity of results to weak coupling regimes.

At the same time, another class of approaches for treating complex non-Markovian dynamics have been developed using the idea of mapping the initial problem onto a simpler Markovian one \cite{Lambert2019,Pleasance2020,Tamascelli2018,Mascherpa2020,Garraway1997,Dalton2001,Mazzola2009,Dalton2003,Garraway2006,IlesSmith2014,IlesSmith2016,Diosi2012,Strasberg2016,Imamoglu1994,Stenius1996,Breuer2004,Woods2014,Martinazzo2011,Burghardt2012,Hughes2009a,Hughes2009b,Arrigoni2013,Dorda2014,Chen2019}. In particular, the pseudomode method \cite{Pleasance2020,Garraway1997,Dalton2001,Mazzola2009,Dalton2003,Garraway2006} implements such a mapping by expanding the system with a set of discrete bosonic modes (the pseudomodes), whose properties are derived from certain analytical features of the environment spectral density. The enlarged system comprising the original system and pseudomodes obeys an exact Markovian master equation which can then be efficiently simulated using standard techniques \cite{Dalibard1992,Gisin1992,Plenio1998}. Previously, this method has been applied to open system models valid within the rotating wave approximation (RWA) and with zero temperature environments \cite{Pleasance2020,Garraway1997}, and has recently seen extensions to non-RWA cases involving several specific types of spectral density function \cite{Lambert2019,Tamascelli2018}.

In this Letter, we go beyond all previous limitations to extend the validity of the pseudomode mapping to a completely generic class of open system models. We achieve this by adapting a general proof from \cite{Tamascelli2018} to show that the reduced dynamics of an open system coupled to bosonic environment can be reproduced exactly when coupled to a small number of pseudomodes via a possibly non-Hermitian form of interaction. To demonstrate the method, we apply it to a paradigmatic open system model---the spin-boson model---finding it able to capture the exact non-Markovian dynamics at strong system-bath coupling, whilst simultaneously being implemented at low computational costs. 

\textit{Physical and auxiliary models.---} We consider a microscopic model of a generic open system $S$ coupled linearly to a bosonic environment $E$ with total Hamiltonian $H = H_S + H_E + H_I$. Here, $H_S$ and $H_E$ are the free Hamiltonians of the system and environment; the latter is written as $H_E = \sum_k\omega_ka^{\dagger}_ka_k$, where $\omega_{k}$ is the frequency of the $k$-mode of the environment and $a_{k}$ ($a^{\dagger}_{k}$) is the corresponding annihilation (creation) operator satisfying $[a_{k},a^{\dagger}_{k'}]=\delta_{kk'}$ ($\hbar=1$). We assume an interaction of the form $H_I = A\otimes B$, with $A=A^{\dagger}$ and $B=\sum_kg_k(a_k+a^{\dagger}_k)$ observables of the system and environment, respectively. Note that the more general case with $H_I$ replaced by $H_I=\sum_{\alpha}A_{\alpha}\otimes B_{\alpha}$ could also be considered, but we choose to avoid it for simplicity. \\
\indent Our main focus here will be on describing the non-Markovian dynamics of open quantum systems interacting with environments that are both Gaussian and stationary---namely, environments which satisfy $\commutator{H_E}{\rho_E(0)}=0$, where $\rho_E(0)$ is a Gaussian state of $E$. Consequently, if the initial state of the total system $S+E$ is factorized as $\rho_{SE}(0)=\rho_S(0)\otimes\rho_E(0)$, then the reduced system dynamics obtained via
\begin{equation}\label{eq:rho_S}
	\rho_S(t) = \text{Tr}_E\Big\{e^{-iHt}\rho_S(0)\otimes\rho_E(0)e^{iHt}\Big\},
\end{equation}
will be fully determined by the two-time correlation function (assuming vanishing first order moments $\langle B(t)\rangle = 0$)
\begin{align}\label{eq:TTCF_def}
	C(\tau) &= \langle B(t+\tau)B(t)\rangle \nonumber\\
		    &= \frac{1}{2\pi}\int^{\infty}_{-\infty}d\omega\,\gamma(\omega)e^{-i\omega\tau}, \qquad \tau\geq0,
\end{align}
with $\langle B(t+\tau)B(t)\rangle = \text{Tr}_E[B(\tau)B\rho_E(0)]$ and $B(\tau) = e^{iH_E\tau}Be^{-iH_E\tau}$. Here we have also introduced the spectral density (SD) $\gamma(\omega)$ as the Fourier transform of the correlation function $C(\tau)$. By definition, the SD is a positive, real-valued function describing the frequency dependent properties of the coupling between the system and environment \cite{BreuerTOQS2002}. 

In the following it will prove very useful to model the interaction using a generalized form of spectral density. More precisely, the only assumptions we shall impose on $\gamma(\omega)$ are for it to be a meromorphic function when analytically continued to the complex $\omega$-plane, and for $\gamma(\omega)$ to fall of faster than $\sim O(1/|\omega|)$ in the limit $|\omega|\rightarrow\infty$. The usefulness of these assumptions is that they allow us to evaluate the two-time correlation function (\ref{eq:TTCF_def}) analytically in terms of the poles and residues of $\gamma(\omega)$ in the lower-half complex $\omega$-plane. To do so, let us first enumerate these poles according to the index $l=1,...,N$. By then writing Eq. (\ref{eq:TTCF_def}) as a contour integral $\int d\omega\rightarrow \oint_Cd\omega$, with $C$ a closed semicircular contour in lower-half plane, we can straightforwardly apply the residue theorem to obtain 
\begin{equation}\label{eq:TTCF_phys}
	C(\tau) = -i\sum^{N}_{l=1}r_le^{-iz_l\tau},\qquad \tau\geq 0,
\end{equation}
where $z_l = \xi_l - i\lambda_l$ are the locations of the poles of $\gamma(\omega)$, and $r_l=\text{Res}_{\omega=z_l}[\gamma(\omega)]$ are their corresponding residues. By setting $\tau=0$ in Eqs. (\ref{eq:TTCF_def}) and (\ref{eq:TTCF_phys}) we may also establish the relation $\sum_lr_l = -(2\pi i)^{-1}\int d\omega\,\gamma(\omega)$, implying the residues $r_l$ have no net real part. 

\begin{figure}[t!]
	\centering
	\includegraphics[width=.46\textwidth]{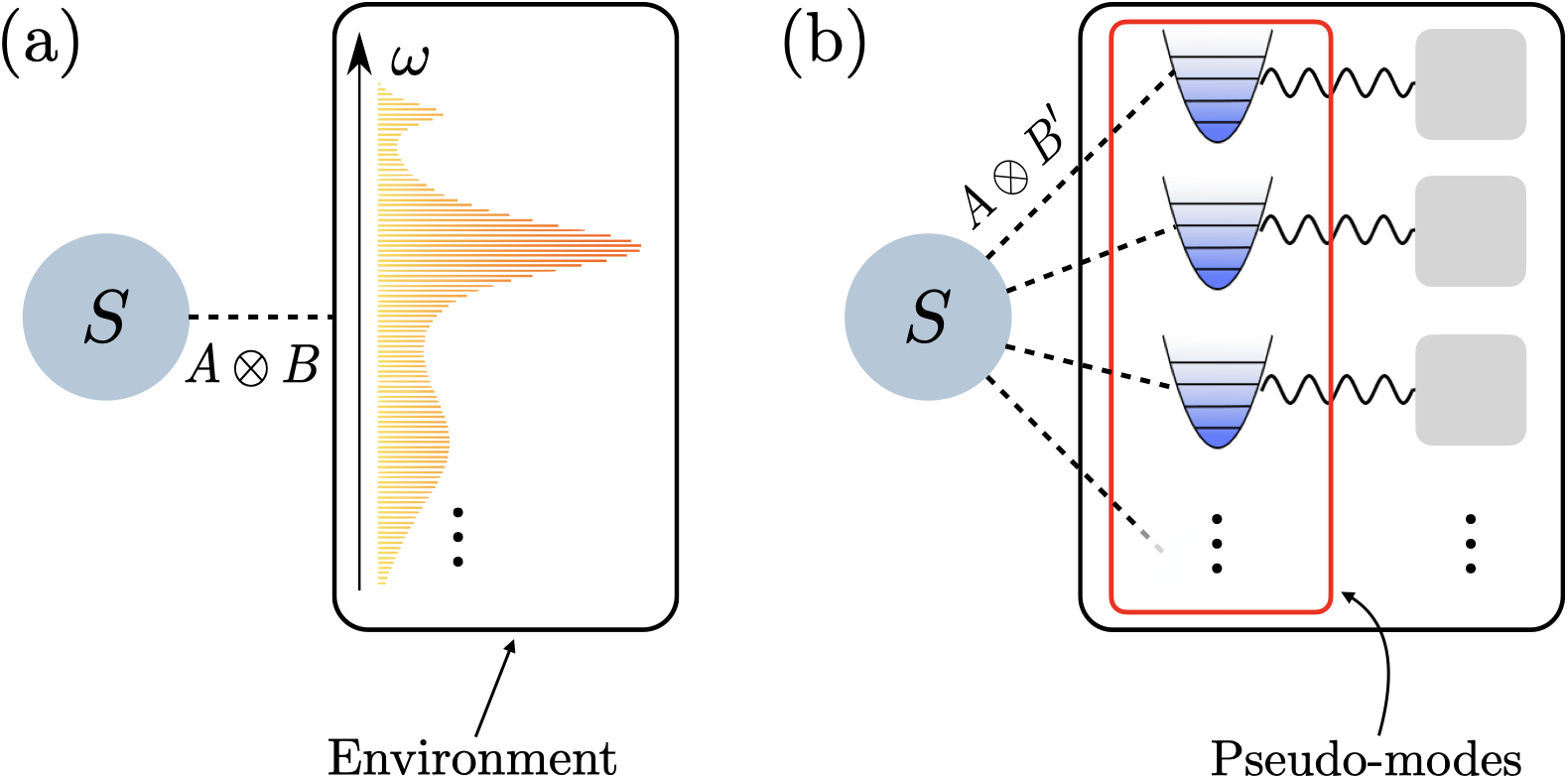}
	\caption{\label{fig:1} Schematic representation of the physical and auxiliary models. (a) An open system $S$ couples to a structured environment $E$ with a frequency dependent SD. (b) The same system couples to a finite number of pseudomodes $M$ with non-unitary evolution, whose dynamics when attached to $S$ is governed by the Markovian master equation (\ref{eq:ME}).}
\end{figure}

We now proceed to outline the auxiliary model that will allow us to represent the exact time evolution of $\rho_S(t)$ within an enlarged Markovian open system (see Fig. \ref{fig:1}). Here, the physical environment $E$ is replaced with a set of discrete bosonic modes---the pseudomodes $M$---which are defined in terms of the positions and residues of the poles of the SD (\ref{eq:TTCF_def}). The Hamiltonian of the enlarged system is given as $H_0 = H_S + H_M + H'_I$, whereby
\begin{align}
	H_M &= \sum_l\xi_lb^{\dagger}_lb_l, \\ 
	H'_I &= A\otimes B'. \label{eq:H'_I}
\end{align}
Above, $b_l$ ($b^{\dagger}_l$) is the annihilation (creation) operator of the $l$-pseudomode satisfying the bosonic commutation relation $[b_l,b^{\dagger}_{l'}]=\delta_{ll'}$, and 
\begin{equation}\label{eq:B'}
	B' = \sum_lg'_l\Big(b_l + b^{\dagger}_l\Big), \qquad g'_l = \sqrt{-ir_l}.
\end{equation}
Notice here that the coupling constant $g'_l$ only appears outside the bracket rather than $g'_l$ and its complex conjugate; as such, the Hamiltonian $H_0$ will generally be \textit{non-Hermitian} due to the residues $r_l$ on which the couplings $g'_l$ depend being complex, except for cases in which the SD adopts a Lorentzian structure. The implications of this will be discussed in more detail below once we have introduced the main result. \\
\indent For the current configuration, the combined density matrix $\rho(t)$ of the open system and pseudomodes obeys a Markovian master equation 
\begin{equation}\label{eq:ME}
	\frac{d}{dt}\rho(t) = \mathcal{L}\rho(t) = -i[H_0,\rho(t)] + \mathcal{D}\rho(t),
\end{equation}
with the superoperator
\begin{equation}
	\mathcal{D}\rho = 2\sum_l\lambda_l\bigg(b_l\rho b^{\dagger}_l - \frac{1}{2}\big\{b^{\dagger}_lb_l,\rho\big\}\bigg),
\end{equation}
describing the local dissipation of each pseudomode occurring at rate $2\lambda_l$. The reduced system state evolving under (\ref{eq:ME}) can then be obtained as
\begin{equation}\label{eq:rho'_S}
	\rho'_S(t) = \text{Tr}_M\Big\{e^{\mathcal{L}t}\rho_S(0)\otimes\rho_M(0)\Big\},
\end{equation}
where we have assumed the pseudomodes to be initially uncorrelated with the system. 

With these details in place, we now look to establish an exact equivalence of the reduced system dynamics expressed through Eqs. (\ref{eq:rho_S}) and (\ref{eq:rho'_S}). We proceed by utilizing an approach recently developed in \cite{Tamascelli2018}, in which the authors introduced a rigorous proof for determining the equivalence of an open system dynamics resulting from two different types of environment; one comprising a Gaussian bosonic reservoir with free unitary evolution, and the other a set of discrete bosonic modes, that together with the open system satisfy a Lindblad-type evolution. In the Supplemental Material we adapt this proof in accordance with \cite{Pleasance2020,Lambert2019} to account for the non-Hermitian form of coupling between the system and pseudomodes \footnote{See Supplemental Material at [URL will be inserted by publisher] for further details on the proof.}. The proof is performed in two stages: first, by purifying $\rho(t)$ on an extended state space $S+M+R$, we show that the reduced dynamics (\ref{eq:rho'_S}) can be exactly reproduced in an extended auxiliary model where the pseudomodes are individually coupled to Markovian reservoirs $R$, and the total state of $S+M+R$ follows a \textit{pseudo-unitary} evolution. Second, we show that the reduced system dynamics in the physical model is equivalent to the reduced dynamics of $S$ in the extended auxiliary model assuming the auxiliary environment $M+R$ is initially in a Gaussian state. Since the free evolution of both environments $E$ and $M+R$ is (pseudo-) unitary, this is achieved by matching the two-time correlation functions $C(\tau)$ and $C'(\tau)$ of the respective coupling operators $B$ and $B'$. Overall, we then find that
\begin{equation}\label{eq:equiv}
	\rho'_S(t) = \rho_S(t). 
\end{equation}
Establishing an exact equivalence between the reduced dynamics of $S$ in the two models means we have not only reaffirmed the results of \cite{Lambert2019,Pleasance2020,Garraway1997}, but also now extended them to a generic class of open system models within the restrictions above. 

\textit{Application to arbitrary SDs.---} Following on from this proof, we will now discuss some details relevant to the application of the result. One key consideration is that because the correspondence in Eq. (\ref{eq:equiv}) is entirely determined by the equivalence between the two-time correlation functions $C(\tau)$ and $C'(\tau)$, the mapping will only be exact when $C(\tau)$ can be written as a finite sum of complex exponentials. This is equivalent to saying $\gamma(\omega)$ is constrained to have the same frequency dependence as the effective spectral density $\gamma'(\omega)=\int^{\infty}_{-\infty}d\tau\,C'(\tau)e^{i\omega\tau}$:
\begin{equation}\label{eq:gm_Lorz}
	\gamma'(\omega) = 2\sum^N_{l=1}\frac{r^R_l(\omega-\xi_l) + r^I_l\lambda_l}{(\omega-\xi_l)^2+\lambda^2_l},
\end{equation}
where $r_l \equiv r^R_l + ir^I_l$. For other forms of SD, including those related to thermal baths (see below), the correspondence between $\gamma(\omega)$ and $\gamma'(\omega)$ cannot be made exact \cite{Tamascelli2018,Mascherpa2020}, and so for these cases we will instead rely on optimizing the parameters of Eq. (\ref{eq:gm_Lorz}) as those most accurately representing the original SD function to obtain 
\begin{equation}
	C(\tau)\approx C'(\tau), \quad \tau\geq0. 
\end{equation}
In practice this can be achieved by fitting the SD of the physical problem with a set of $N$ basis functions $\gamma'_l(\omega)=2[r^R_l(\omega-\xi_l)+r^I_l\lambda_l]/[(\omega-\xi_l)^2+\lambda^2_l]$---or with $N$ complex exponentials $C'_l(\tau) = -ir_l\exp(-iz_l\tau)$ in the time domain---where each pole of $\gamma'_l(\omega)$ in the lower-half plane is associated to one of the $N$ pseudomodes in the auxiliary model. The residues of these poles are again used to parameterize the couplings $g'_l$ within the enlarged system Hamiltonian. In fact, since the non-Hermitian terms of $H_0$ are generated through the imaginary part of these couplings $g'_l$, their role in determining $\gamma'(\omega)$ may be conveniently illustrated by imposing a Hermitian form of interaction, i.e., by setting $r^R_l=0$ and $r^I_l>0$, so that $g'_l$ are constrained to be real. From Eq. (\ref{eq:gm_Lorz}) we see this corresponds to $\gamma'(\omega)$ being written as a sum of positively weighted Lorentzians, implying the effect of the non-Hermitian terms is to generalize the form of $\gamma'(\omega)$ from a simple sum of Lorentzians to more complicated functions which are either (i) non-Lorentzian, or (ii) Lorentzian but with negative weights. As one might expect, these functions are capable of representing a much greater variety of environments compared to when $g'_l$ are real \cite{Imamoglu1994,Stenius1996}, thus enabling a more efficient simulation of the problem in general. 

\begin{figure*}[t!]
	\centering
	\includegraphics[width=\textwidth]{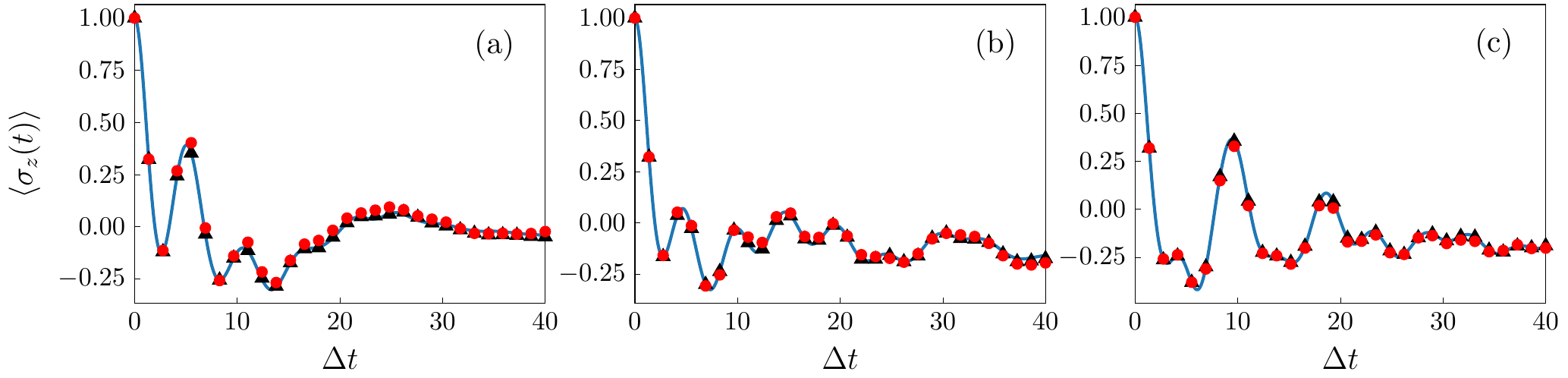}
	\caption{\label{fig:2} Time evolution of $\langle\sigma_z(t)\rangle$ obtained from the pseudomode master equation (solid blue line and solid black triangles) and the HEOM (solid red points), (a) for $\omega_0=0.25\Delta$, (b) $\omega_0=0.5\Delta$, and (c) $\omega_0=\Delta$. The black triangles indicate the result generated by mapping the $n=1$ term of Eq. (\ref{eq:M(t)}) onto a single pseudomode and with a local dephasing term acting on the TLS. The other parameters are $\beta\Delta = 1$, $\epsilon=0.5\Delta$, $\Gamma=0.05\Delta$ and $\alpha=0.25\Delta$. For all plots, the Matsubara part of the correlation function (\ref{eq:M(t)}) is evaluated up to a total of $1.5\times 10^3$ terms.}
\end{figure*}

We finish this section by briefly commenting on the connection of our approach with that of Mascherpa \textit{et. al.} in Ref. \cite{Mascherpa2020}. There, the authors have introduced an analogous treatment based on using network of damped harmonic oscillators to replicate the effect of a bosonic environment on an open system, assuming the bath correlation function can be approximated as a sum of complex exponentials (\ref{eq:TTCF_phys}). Our approach differs from \cite{Mascherpa2020} primarily in the ansatz used to construct the effective environment as well as the master equation used to model the dynamics of the enlarged system; more precisely, the auxiliary modes used to replace the environment in \cite{Mascherpa2020} are interacting, and their dynamics when coupled to the system is described by a Lindblad master equation. The main consequence of these differences is that the couplings and decay rates of the auxiliary modes cannot be directly related to those of the correlation function (\ref{eq:TTCF_phys}), and so these parameters must be extracted by performing an additional step of inverting a system of non-linear equations depending on the weights and exponents of $C'(\tau)$. In this regard, our method may offer certain numerical advantages for situations where it is difficult to perform such an inversion, since apart from the fitting of the two-time correlation function $C(\tau)$, no further steps are necessary in order to simulate the physical problem.

\textit{Spin-boson model.---} We now proceed to outline an example of the mapping applied to the thermal bath case, where the dynamics of the system is described by the spin-boson Hamiltonian \cite{Leggett1987}
\begin{equation}
	H = \frac{\epsilon}{2}\sigma_z + \frac{\Delta}{2}\sigma_x + \sum_{k}\omega_{k}a^{\dagger}_{k}a_{k} + \sigma_z\otimes\sum_{k}g_{k}\big(a_{k} + a^{\dagger}_{k}\big). 
\end{equation}
Here, the system and interaction Hamiltonians $H_S=(\epsilon/2)\sigma_z + (\Delta/2)\sigma_x$ and $H_I=\sigma_z\otimes B$ are written in terms of the Pauli spin-$\frac{1}{2}$ operators $\sigma_i$ ($i=x,y,z$) in the subspace $\{|g\rangle,|e\rangle\}$ of the two-level system $S$ (TLS), with $\sigma_z = |e\rangle\langle e| - |g\rangle\langle g|$. The environment is initially in a thermal equilibrium state $\rho_E(0) = \exp(-\beta H_E)/\text{Tr}\big[\exp(-\beta H_E)\big]$ at temperature $T=1/\beta$ ($k_B=1$), such that the corresponding two-point correlation function reads 
\begin{equation}\label{eq:TTCF}
	C(\tau) = \frac{1}{\pi}\int^{\infty}_0d\omega\,J(\omega)\Big[\coth\left(\frac{\beta\omega}{2}\right)\cos(\omega\tau) -i\sin(\omega\tau)\Big],
\end{equation}
having introduced the one-sided spectral density function $J(\omega)=\pi\sum_{k}g^2_{k}\delta(\omega-\omega_{k})$ \cite{BreuerTOQS2002}. For convenience, we will consider $J(\omega)$ to be in an underdamped Brownian oscillator form $J(\omega)=\alpha\omega^2_0\Gamma\omega/[(\omega^2_0-\omega^2)^2 + \Gamma^2\omega^2]$, where $\Gamma$ and $\omega_0$ define a characteristic width and resonance frequency, respectively, and $\alpha$ is the reorganization energy of the bath \cite{Garg1985}. 

The correlation function corresponding to $\gamma(\omega)$ can be determined analytically as $C(\tau) = C_0(\tau) + M(\tau)$, where 
\begin{align}\label{eq:C_0}
	&C_0(\tau) = \frac{\alpha\omega^2_0}{4\Omega}e^{-\Gamma\tau/2}\left[\coth\left(\frac{\beta(\Omega-i\Gamma/2)}{2}\right)e^{i\Omega\tau} + \text{h.c.}\right] \nonumber\\
	&\quad + \frac{\alpha\omega^2_0}{4\Omega}e^{-\Gamma\tau/2}\left[e^{-i\Omega\tau} - e^{i\Omega\tau}\right],
\end{align}
and
\begin{align}\label{eq:M(t)}
	&M(\tau) = \sum^{\infty}_{n=1}c_ne^{-\nu_n\tau}, \\
	&c_n = \frac{-2\alpha\Gamma\omega^2_0\nu_n}{\beta[\nu^2_n + (\Omega + i\Gamma/2)^2][\nu^2_n + (\Omega - i\Gamma/2)^2]}, \nonumber
\end{align}
with $\Omega = \sqrt{\omega^2_0 - (\Gamma/2)^2}$, and Matsubara frequencies $\nu_n = 2\pi n/\beta$. Since the analytic part of the correlation function (\ref{eq:C_0}) is already in form of Eq. (\ref{eq:TTCF_phys}), we choose to capture the effect of the remaining Matsubara contributions by first fitting $M(\tau)$ with the sum of two real exponentials \cite{Lambert2019}, such that $M'(\tau) = W_1e^{-\gamma_1\tau} + W_2e^{-\gamma_2\tau}$, where $C'(\tau) = C_0(\tau) + M'(\tau)$. We emphasize that it is the use of complex couplings $g'_l$ which allows us to represent $C_0(\tau)$ exactly using only two pseudomodes. In Fig. \ref{fig:2}, we compare the reduced system dynamics computed from the pseudomode master equation and the HEOM for different values of $\omega_0$ in the strong coupling and narrow bath regime, assuming the TLS to be initialized in the state $\rho_S(0)=|e\rangle\langle e|$. The two approaches have been implemented numerically using the Python library QuTiP \cite{Johansson2013} and the integrated QuTiP package BoFiN-HEOM provided in \cite{Lambert2020,Bofin2020}. As expected, we find close correspondence between the predictions of the master equation and HEOM over both short and long time scales. In addition to the fitting approach, we also choose to simulate the TLS dynamics by retaining only a single Matsubara term in the expansion (\ref{eq:M(t)}), while treating the remaining terms as delta functions \cite{Ishizaki2005}. Since $M(\tau)$ is sharply peaked around $\tau=0$, for $\tau\ll 1/\nu_n$ one can write it approximately as $M(\tau) \approx c_1e^{-\nu_1\tau} + \sum^{\infty}_{n=2}\frac{c_n}{\nu_n}\delta(\tau)$. The first term of $M(\tau)$ is then mapped to a single pseudomode following the same procedure outlined above. On the other hand, the delta function can be accounted for by adding a local dephasing term to the Liouvillian in Eq. (\ref{eq:ME}) of the form $\mathcal{L}_{D}\rho = 2\gamma_D(\sigma_z\rho\sigma_z - \rho)$, where $\gamma_D=\sum_{n>1}(c_n/\nu_n)$. This is done analogously to how the `Taninmura terminator' is implemented within the HEOM. For the cases shown (see Fig. \ref{fig:2}), we were able with the approximated expansion to simulate the master equation at an even smaller cost than using the full Matsubara decomposition (\ref{eq:M(t)}), which becomes particularly advantageous at strong coupling when the local dimension of each pseudomode is generally needed to be higher for convergence. 

\textit{Summary \& discussion.---} In conclusion, we have presented a novel framework for simulating the non-equilibrium behavior of open quantum systems within structured environments. Our method relies on replicating the effect of a bosonic environment on a general open system with a collection of pseudomodes, i.e., discrete bosonic modes, where each pseudomode is connected to one of the poles of the SD in the lower-half complex frequency plane. Since the mapping is non-perturbative in the original system-bath coupling, our formalism enables the full inclusion of memory effects within the reduced dynamics of the simulated problem. This has been demonstrated for the spin-boson model by benchmarking the results obtained from the pseudomode master equation against those from the numerically exact HEOM. Besides having numerous theoretical applications, our findings open up ways to possibly analyze the complex dynamics occurring within experimental setups relevant to e.g., quantum thermodynamics \cite{Strasberg2016,Newman2017}, trapped ions \cite{Lemmer2018} and quantum many-body physics \cite{Daley2014}. Future developments of the method could look at extensions to fermionic environments \cite{Arrigoni2013,Dorda2014,Chen2019}, as well as making improvements to its performance based on the accuracy of the fitting algorithm. 

\textit{Acknowledgements.---} This work is based upon research supported by the South African Research Chair Initiative, Grant No. 64812 of the Department of Science and Innovation and the National Research Foundation of the Republic of South Africa. Support from the NICIS (National Integrated Cyber Infrastructure System) e-research grant QICSI7 is kindly acknowledged.

\bibliographystyle{apsrev4-2} 

\bibliography{PM_PRL.bib}

\appendix
\widetext
\pagebreak

\section*{Supplemental material}\label{appenA}

In this Supplemental Material we prove the general equivalence between the open system dynamics derived in both the physical and auxiliary models discussed in the main text. To do this, we specially modify the proof in \cite{Tamascelli2018} to account for the non-Hermitian form of coupling between the open system and pseudomodes in line with Refs. \cite{Pleasance2020,Lambert2019}. In particular, we note that while our strategy is closely aligned to one presented by Lambert \textit{et. al.} in \cite{Lambert2019}, their result was only formally derived for a specific form of SD and zero-temperature environment. Here we not only extend this result to arbitrary Gaussian states (including e.g., thermal states), but also to a generalized class of SDs satisfying the analyticity constraints above. 

\subsection{Auxiliary and extended auxiliary models}

First, we are going to prove that the reduced dynamics stemming from the master equation (\ref{eq:ME}) in the main text is equivalent to the reduced system dynamics of an extended auxiliary model $S+M+R$, where each of the pseudomodes $M$ is coupled to a local Markovian reservoir. The total Hamiltonian of this extended model reads
\begin{equation}
	H' = H_S + H_{E'} + H'_I
\end{equation} 
where $H_{E'} = H_M + H_R + V_{MR}$,
\begin{align}
	&H_R =  \sum_l\int^{\infty}_{-\infty}d\omega\,\omega a^{\dagger}_{Rl}(\omega)a_{Rl}(\omega),\label{eq:H_M} \\
	&V_{MR} = \sum_l\int^{\infty}_{-\infty}\sqrt{\frac{\lambda_l}{\pi}}d\omega\bigg(b^{\dagger}_la_{Rl}(\omega)+\text{h.c.}\bigg),  \label{eq:V_MR}
\end{align}
with $a_{Rl}(\omega)$ ($a^{\dagger}_{Rl}(\omega)$) the annihilation (creation) operator for excitation of frequency $\omega$ in the $l$-reservoir of $R$, and $[a_{Rl}(\omega),a^{\dagger}_{Rl'}(\omega')]=\delta_{ll'}\delta(\omega-\omega')$. Here, $H_R$ is the free Hamiltonian of the reservoirs, while $V_{MR}$ describes the total $M+R$ interaction. Note that the frequency independence of the coupling constants in Eq. (\ref{eq:V_MR}) ensures the reservoirs each have a vanishing correlation time in accordance with \cite{Gardiner2005}.

For a non-unitary evolution $S+M$ in the auxiliary model, 
\begin{equation}\label{eq:rho}
	\rho(t) = e^{\mathcal{L}t}\Big[\rho_S(0)\otimes\rho_M(0)\Big],
\end{equation}
we now proceed to show the density matrix 
\begin{equation}\label{eq:rho_SMR}
	\rho_{SMR}(t) = e^{-iH't}\big[\rho(0)\otimes|0\rangle\langle0|_R\big]e^{iH't}
\end{equation}
yields a valid purification of $\rho(t)$ on the extended state space $S+M+R$, such that $\rho(t) = \text{Tr}_R[\rho_{SMR}(t)]$ (here, $|0\rangle_R=\bigotimes^N_{l=1}|0_l\rangle$ denotes the collective vacuum state of the reservoirs). In particular, it is worth noting that since the Hamiltonian $H'$ is in principle non-Hermitian, Eq. (\ref{eq:rho_SMR}) defines a pseudo-unitary evolution which neither preserves Hermiticity or positivity of the initial state $\rho_{SMR}(0)$. The `pseudo' prefix refers to the fact that the righthand side of (\ref{eq:rho_SMR}) is acted on by the inverse of the time evolution operator $\exp(-iH't)$, rather than its Hermitian conjugate, as might otherwise be employed for a time evolution involving a non-Hermitian generator. 

Using Eq. (\ref{eq:rho_SMR}), the Heisenberg equation of motion for an arbitrary operator $O=O^{\dagger}$ of the enlarged system is given as ($O(t)=e^{iH't}Oe^{-iH't}$)
\begin{align}\label{eq:HE_O}
	&\frac{d}{dt}O(t) = -i[O(t),H_0(t)]  \nonumber\\
	&\quad-i\sum_l\int d\omega\sqrt{\frac{\lambda_l}{\pi}}\Big\{[O(t),b^{\dagger}_l(t)]a_{Rl}(\omega,t) + a^{\dagger}_{Rl}(\omega,t)[O(t),b_l(t)]\Big\}.
\end{align}
Eliminating the dependence on the reservoir operators $a_{Rl}(\omega,t)$ and $a^{\dagger}_{Rl}(\omega,t)$ subsequently yields the quantum Langevin equation
\begin{align}\label{eq:QLE}
	\frac{d}{dt}O(t) &= -i[O(t),H_0(t)] + \sum_l\lambda_l\left\{b^{\dagger}_l(t)[O(t),b_l(t)] - [O(t),b^{\dagger}_l(t)]b_l(t)\right\} \nonumber\\
			       &-i\sum_l\sqrt{2\lambda_l}\left\{[O(t),b^{\dagger}_l(t)]a^{(in)}_{Rl}(t) + a^{(in)\dagger}_{Rl}(t)[O(t),b_l(t)]\right\},
\end{align}
where
\begin{equation}
	a^{(in)}_{Rl}(t) =\frac{1}{\sqrt{2\pi}}\int^{\infty}_{-\infty}d\omega\,e^{-i\omega t}a_{Rl}(\omega). 
\end{equation}
obeys the quantum white noise relation \cite{Gardiner2005}
\begin{equation}
	[a^{(in)}_{Rl}(t), a^{(in)\dagger}_{Rl'}(s)] = \delta_{ll'}\delta(t-s). 
\end{equation}
Next we can write down the expectation value $\langle O(t)\rangle = \text{Tr}\big[O(t)\rho_{SMR}(0)\big]$ of Eq. (\ref{eq:QLE}) for an initially factorized density matrix $\rho_{SMR}(0)=\rho(0)\otimes|0\rangle\langle 0|_R$. In this case, the second line of Eq. (\ref{eq:QLE}) vanishes from
\begin{equation}
	a^{(in)}_{Rl}(t)\rho_R(0) = 0 = \rho_R(0)a^{(in)\dagger}_{Rl}(t),
\end{equation}
and so 
\begin{align}\label{eq:QLE_ave}
	&\frac{d}{dt}\langle O(t)\rangle = -i\text{Tr}\Big[\commutator{O(t)}{H_0(t)}\rho_{SMR}(0)\Big] \nonumber\\
	&\quad + \sum_l\lambda_l\,\text{Tr}\left[\left\{b^{\dagger}_l(t)[O(t),b_l(t)] - [O(t),b^{\dagger}_l(t)]b_l(t)\right\}\rho_{SMR}(0)\right].
\end{align}
Because Eq. (\ref{eq:QLE_ave}) only contains operators acting on the enlarged system Hilbert space, we are then able to use the cyclic property of the trace 
\begin{equation}
	\langle O(t)\rangle \equiv \text{Tr}[O(t)\rho_{SMR}(0)] = \text{Tr}[O\rho'(t)]
\end{equation}
to obtain
\begin{equation}\label{eq:ME_appen}
	\frac{d}{dt}\rho'(t) = -i[H_0,\rho'(t)] + 2\sum_l\lambda_l\left(b_l\rho'(t)b^{\dagger}_l - \frac{1}{2}\big\{b^{\dagger}_lb_l,\rho'(t)\big\}\right),
\end{equation}
which, from being identical to the master equation (\ref{eq:ME}) in the main text, and with $\rho'(t)$ subject to the same initial condition as $\rho(t)$ [Eq. (\ref{eq:rho})], implies $\rho(t)=\rho'(t)$. Finally, tracing out the pseudomode degrees of freedom from both sides of this expression leads to 
\begin{equation}\label{eq:rho'_S_ext_aux}
	\rho'_S(t) = \text{Tr}_{MR}[\rho_{SMR}(t)]. 
\end{equation}

\subsection{Physical and extended auxiliary models} 

We now move on to prove that the reduced system evolution in the physical model $S+E$ is equivalent to reduced dynamics of $S$ in the extended auxiliary model. As we may recall from Ref. \cite{Tamascelli2018} (and as a direct consequence of Wick's theorem), a sufficient condition to guarantee the equivalence of the open system dynamics evolving under two physically different Gaussian environments is for the moments of their free bath coupling operators to match to second-order. More recently this condition was shown to hold even if coupling between the system and one of the environments is non-Hermitian \cite{Lambert2019,Pleasance2020}, so long as the total density matrix of the system-plus-auxiliary environment obeys a pseudo-unitary evolution as per Eq. (\ref{eq:rho_SMR}). As such, we are able to employ the same arguments here to prove the reduced system density matrices $\rho_S(t)$ and $\text{Tr}_{MR}[\rho_{SMR}(t)]$ share an equal time dependence. To this end, let us assume the auxiliary environment $M+R$ is initially in the vacuum state 
\begin{equation}\label{eq:rho_MR}
	\rho_{MR}(0) = |0\rangle\langle 0|_M\otimes|0\rangle\langle0|_R,
\end{equation}
such that
\begin{equation}
	[H_{E'},\rho_{MR}(0)]=0.
\end{equation}	
The reduced system dynamics of the extended auxiliary model will then be fully determined by the second-order correlation function (assuming vanishing first order moments $\langle B'(t)\rangle=0$) 
\begin{equation}\label{eq:TTCF_aux_def}
	C'(\tau) = \text{Tr}\Big[B'(\tau)B'(0)\rho_{MR}(0)\Big], \qquad \tau\geq 0,
\end{equation}
where $B'(\tau) = e^{iH_{E'}\tau}B'e^{-iH_{E'}\tau}$. 

By now inserting the explicit expression for $B'(\tau)$ into the above, we obtain 
\begin{equation}\label{eq:TTCF_aux_def_2}
	C'(\tau) = -i\sum^N_{l,m=1}\sqrt{r_lr_m}\langle b_l(\tau)b^{\dagger}_m(0)\rangle,
\end{equation}
given that all second-order moments involving normal ordered products of operators $b_l$ and $b^{\dagger}_l$ (e.g., $\langle b^{\dagger}_l(\tau)b_m(0)\rangle$) are zero for the initial state (\ref{eq:rho_MR}). The time dependence of the correlation functions $\langle b_l(\tau)b^{\dagger}_m(0)\rangle$ is contained in free evolution of the pseudomodes, whose Heisenberg equation of motion reads 
\begin{equation}
	\frac{d}{d\tau}b_l(\tau) = -i[b_l(\tau),H_{E'}] = -i\xi_lb_l(\tau) - i\sqrt{\frac{\lambda_l}{\pi}}a_{Rl}(\omega,t).
\end{equation}
After eliminating the reservoir variables $a_{Rl}(\omega,\tau)$ this equation becomes
\begin{equation}
	\frac{d}{d\tau}b_l(\tau) = -iz_lb_l(\tau) - i\sqrt{2\lambda_l}a^{(in)}_{Rl}(t),
\end{equation}
which can be formally solved to yield
\begin{equation}
	b_l(\tau) = e^{-iz_l\tau}b_l(0) -i\sqrt{2\lambda_l}\int^{\tau}_0ds\,e^{-iz_l(\tau-s)}a^{(in)}_{Rl}(s). 
\end{equation}
In turn, we may substitute $b_l(\tau)$ into the above and use the initial condition $\langle b_l(0)b^{\dagger}_m(0)\rangle=\delta_{lm}$, together with $\Big\langle b_m(0)a^{(in)}_{Rl}(s)\Big\rangle = 0$, to obtain
\begin{equation}\label{eq:TTCF_aux}
	C'(\tau) = -i\sum^N_{l=1}r_le^{-iz_l\tau}, \quad \tau\geq 0.
\end{equation} 
Since the correlation function $C'(\tau)$ is therefore identical to its counterpart $C(\tau)$ in the physical model, we have from the Gaussian property of $\rho_E(0)$ and $\rho_{MR}(0)$ that
\begin{equation}
	\rho_S(t) = \text{Tr}_{MR}[\rho_{SMR}(t)]. 
\end{equation}
Finally, by direct comparison of this result with Eq. (\ref{eq:rho'_S_ext_aux}), it is possible to establish the exact equivalence of the reduced dynamics occurring in both the physical and auxiliary models,
\begin{equation}\label{eq:rho''_S_1}
	\rho_S(t) = \rho'_S(t),
\end{equation}
as stated in Eq. (\ref{eq:equiv}) of the main text.

As an additional remark, it is worth noting that the same result may equally be derived by utilizing an alternative definition of the two-time correlation function (\ref{eq:TTCF_aux_def}) that is consistent with the quantum regression theorem \cite{Gardiner2005,Carmichael1999}. Since the master equation (\ref{eq:ME_appen}) is Markovian, by setting $H_S=0$ and $H'_I=0$, it is possible to show that $C'(\tau)$ can be re-written in the form (see the Supplemental Material of \cite{Tamascelli2018}),
\begin{equation}\label{eq:TTCF_QRT}
	C'(\tau) = \text{Tr}\Big\{B'e^{\mathcal{L}_M\tau}\Big[B'\rho_M(0)\Big]\Big\}, \qquad \tau\geq 0,
\end{equation}
where $\mathcal{L}_M\rho = -i[H_M,\rho]+\mathcal{D}\rho$. From here the proof follows the same steps outlined above: using that $\rho_M(0)=|0\rangle\langle 0|_M$, the correlation function $C'(\tau)$ can again be reduced to the same form as in Eq. (\ref{eq:TTCF_aux_def_2}) where the time dependence of $b_l(\tau)$ is determined by $b_l(\tau)=\exp\big(\mathcal{L}^{\dagger}_M\tau\big)[b_l(0)]$ (here, $\mathcal{L}^{\dagger}_M$ is defined via the duality relation $\text{Tr}\big[A\mathcal{L}_M(B)\big] = \text{Tr}\big[\mathcal{L}^{\dagger}_M(A)B\big]$). Since $b_l$ obeys the eigenoperator relation $\mathcal{L}_Mb_l = -iz_lb_l$, it is then straightforward to demonstrate that the above definition recovers Eq. (\ref{eq:TTCF_aux}), implying $\rho'_S(t)=\rho_S(t)$.

\end{document}